\documentclass[namedreferences]{solarphysics}

\usepackage[hyperref,optionalrh,natbib]{spr-sola-addons} 
\usepackage{graphicx}        
\usepackage{color}           
\usepackage{breakurl}        

\DeclareRobustCommand{\ion}[2]{%
\relax\ifmmode
\ifx\testbx\f@series
{\mathbf{#1\,\mathsc{#2}}}\else
{\mathrm{#1\,\mathsc{#2}}}\fi
\else\textup{#1\,{\mdseries\textsc{#2}}}%
\fi}
\newcommand{\farcs}{\hbox{$.\!\!^{\prime\prime}$}}
\usepackage{natbib} 
\bibpunct{(}{)}{;}{a}{,}{,}

\begin{document}
\begin{article}

\begin{opening}

\title{High-resolution Observations of H$\alpha$ Spectra with a Subtractive Double Pass}

   \author{C.~\surname{Beck}$^{1}$ \sep R.~\surname{Rezaei}$^{2,3}$ \sep D.~\surname{Prasad Choudhary}$^{4}$ \sep S.~\surname{Gosain}$^1$ \sep A.~\surname{Tritschler}$^1$ \sep R.E.~\surname{Louis}$^5$}
        
   \runningtitle{High-resolution H$\alpha$ Observations with a Subtractive Double Pass}
  \runningauthor{C. Beck \textit{et al.}}

   \institute{$^1$: National Solar Observatory (NSO)\\
     $^2$: Instituto de Astrof\'{\i}sica de Canarias (IAC)\\
     $^3$: Departamento de Astrof{\'i}sica, Universidad de La Laguna\\
     $^4$: Department of Physics \& Astronomy, California State University, Northridge\\
     $^5$: Center of Excellence in Space Sciences India (CESSI), Indian Institute of Science Education and Research Kolkata\\
    }
\begin{abstract}
High-resolution imaging spectroscopy in solar physics has relied on Fabry-P{\'e}rot Interferometers (FPIs) in recent years. FPI systems, however, get technically challenging and expensive for telescopes larger than the 1-m class. A conventional slit spectrograph with a diffraction-limited performance over a large field of view (FOV) can be built at much lower cost and effort. It can be converted to an imaging spectro(polari)meter using the concept of a subtractive double pass (SDP). We demonstrate that an SDP system can reach a similar performance as FPI-based systems with a high spatial and moderate spectral resolution across a FOV of 100$^{\prime\prime} \times$100$^{\prime\prime}$ with a spectral coverage of 1\,nm. We use H$\alpha$ spectra taken with a SDP system at the Dunn Solar Telescope and complementary full-disc data to infer the properties of small-scale superpenumbral filaments. We find that the majority of all filaments end in patches of opposite-polarity fields. The internal fine-structure in the line-core intensity of H$\alpha$ at spatial scales of about 0\farcs5 exceeds that in other parameters such as the line width, indicating small-scale opacity effects in a larger-scale structure with common properties. We conclude that SDP systems are a valid alternative to FPI systems when high spatial resolution and a large FOV are required. They also can reach a cadence that is comparable to that of FPI systems, while providing a much larger spectral range.
\end{abstract}

\end{opening}
\section{Introduction}
To trace the fast dynamics of the solar atmosphere requires observations with a high spatial and spectral resolution, a high cadence and a sufficiently large field of view (FOV). The latter depends to some extent on the type of structure under scrutiny, but usually has to cover at least a few granules with a typical extent of 2--3 Mm each for photospheric studies or a few ten Mms for chromospheric topics. 

These days, the most common type of instruments for such purposes are based on Fabry-P{\'e}rot Interferometers (FPIs), \textit{e.g.}, the \textit{CRisp Imaging Spectrometer} \citep[CRISP;][]{scharmer+etal2008} at the  \textit{Swedish Solar Telescope} \citep{scharmer+etal2003aa}, the  \textit{Interferometric Bidimensional Spectrometer} \citep[IBIS;][]{cavallini2006,reardon+cavallini2008} at the  \textit{Dunn Solar Telescope} \citep[DST;][]{dunn1969}, the  \textit{Triple Etalon SOlar Spectrometer} \citep{kentischer+etal1998,tritschler+etal2002} at the  \textit{German Vacuum Tower Telescope} (VTT), the  \textit{GREGOR Fabry-P{\'e}rot Interferometer} \citep{puschmann+etal2006,puschmann+etal2012b}, or the  \textit{Solar Vector Magnetograph} at the  \textit{Udaipur Solar Observatory} \citep{gosain+etal2006}. 

As FPIs were not easily, or not at all, available a few decades ago, the concept of a subtractive double pass (SDP) was used for imaging spectroscopic observations in the past. It presumably was first described and realized in \citet{oehman1950}, followed later by \citet{stenflo1968}. It became more widely known and used through the pioneering work of \citet{mein+blondel1972} that triggered the development of an upgrade to a multi-channel SDP \citep[MSDP;][see also \citeauthor{lopezariste+etal2011} \citeyear{lopezariste+etal2011}]{mein1991} and the installation of (M)SDP systems at different solar telescopes \citep{mein1995}, \textit{e.g.}, the VTT \citep{mein1991}, the \textit{Observatoire du Pic du Midi}, the Meudon solar tower \citep{mein1977}, the  \textit{T{\'e}lescope H{\'e}liographique pour l’Etude du Magn{\'e}tisme et des Instabilit{\'e}s Solaires} (THEMIS) \citep{mein2002}, and the Wroclaw large coronagraph\footnote{Described in a JOSO publication by Rompolt, B., 1993, according to \citet{mein2002}.}.

(M)SDP observations have been used continuously in the last decades up to today \citep[\textit{e.g.},][this list is not intended to be exhaustive in the least way]{mein2002,schmieder+etal2004,mein+etal2009,schmieder+etal2014,heinzel+etal2015}, but with the advent of FPI-based instruments the SDP concept has not been considered for new instrumentation. The fact that many of the (M)SDP observations were either done before adaptive optics (AO) systems were available or were done at telescopes without AO might have contributed to the disregard. For telescopes with a diameter much larger than 1\,m, it becomes, however, difficult to impossible to procure etalons with a sufficiently large diameter and satisfactory optical performance to build an FPI system with diffraction-limited performance, a large FOV and high spectral resolution. Contrary to that, spectrograph systems with a good performance across a large FOV present no impossible technical odds, and even more important, are comparably much cheaper and much easier to realize and operate, while they have the drawback of usually requiring more space than FPIs.

To investigate the performance of an SDP system with high spatial resolution, we converted the \textit{Horizontal Spectrograph} \citep[HSG;][]{dunn+smartt1991} of the DST into an SDP instrument. Thanks to the real-time AO correction, we could ensure that the input illumination for the observations had a spatial resolution down to the diffraction limit. In this study, we describe the SDP setup at the DST in detail in Section \ref{sdp_realize}. We show multiple examples of the spatial resolution and performance that the HSG SDP can achieve. We investigate the SDP performance and the properties of superpenumbral filaments in SDP and complementary full-disc data in Section \ref{secresults}. Section \ref{summ} summarizes our findings while Section \ref{sec_concl} provides our conclusions. The appendices provide additional technical details and example observations.

\section{Subtractive Double Pass Concept}
Even if the SDP concept has already been described in detail elsewhere \citep{mein+blondel1972,mein1991,mein1995,lopezariste+etal2011}, we would like to introduce it once more for a better understanding of the setup used and its operation as described in the following sections. The main idea behind the SDP concept is the usage of a physical mask in a dispersed spectrum to obtain a quasi-monochromatic image. 

\begin{figure}
\centerline{\resizebox{5cm}{!}{\includegraphics{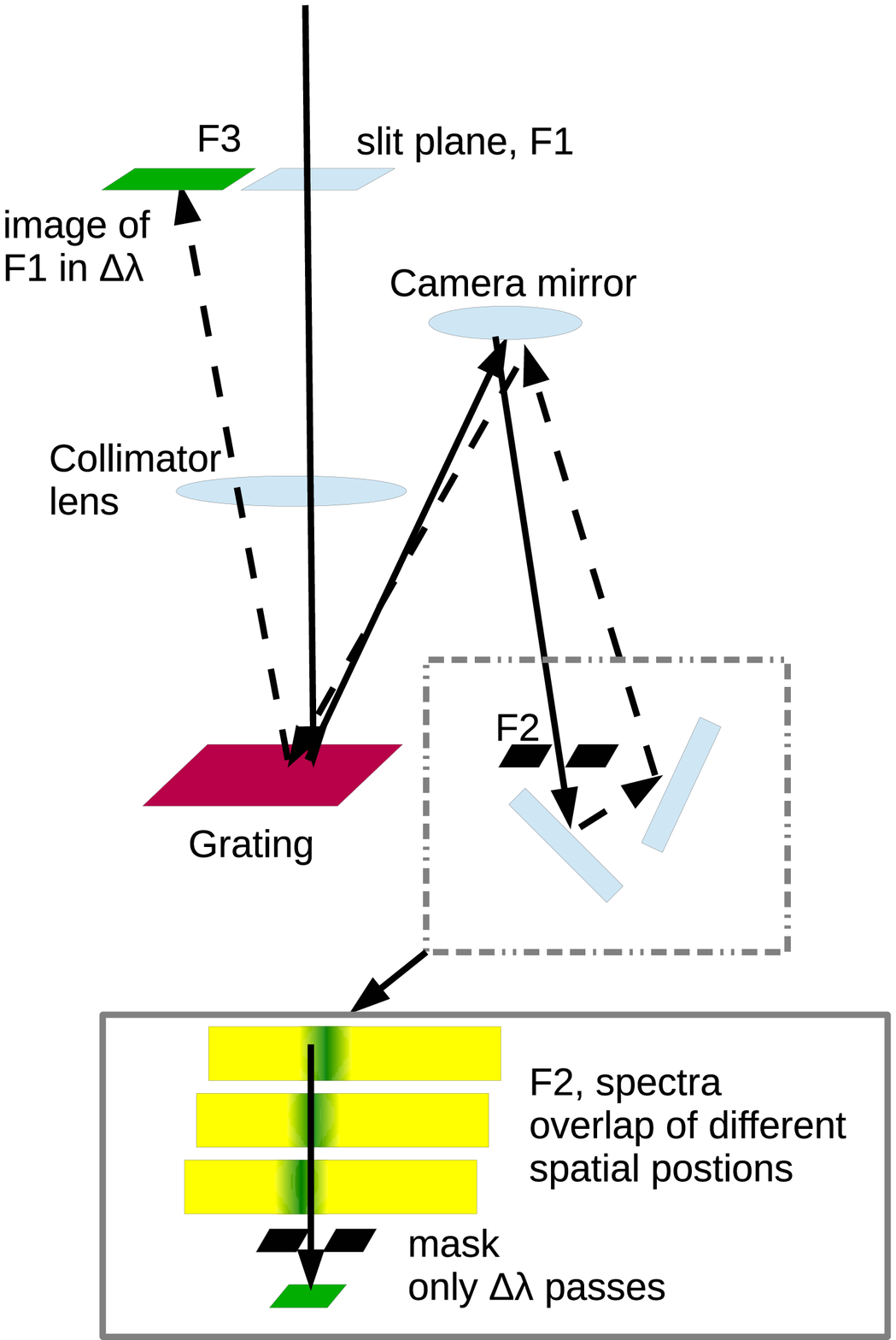}}\resizebox{7cm}{!}{\includegraphics{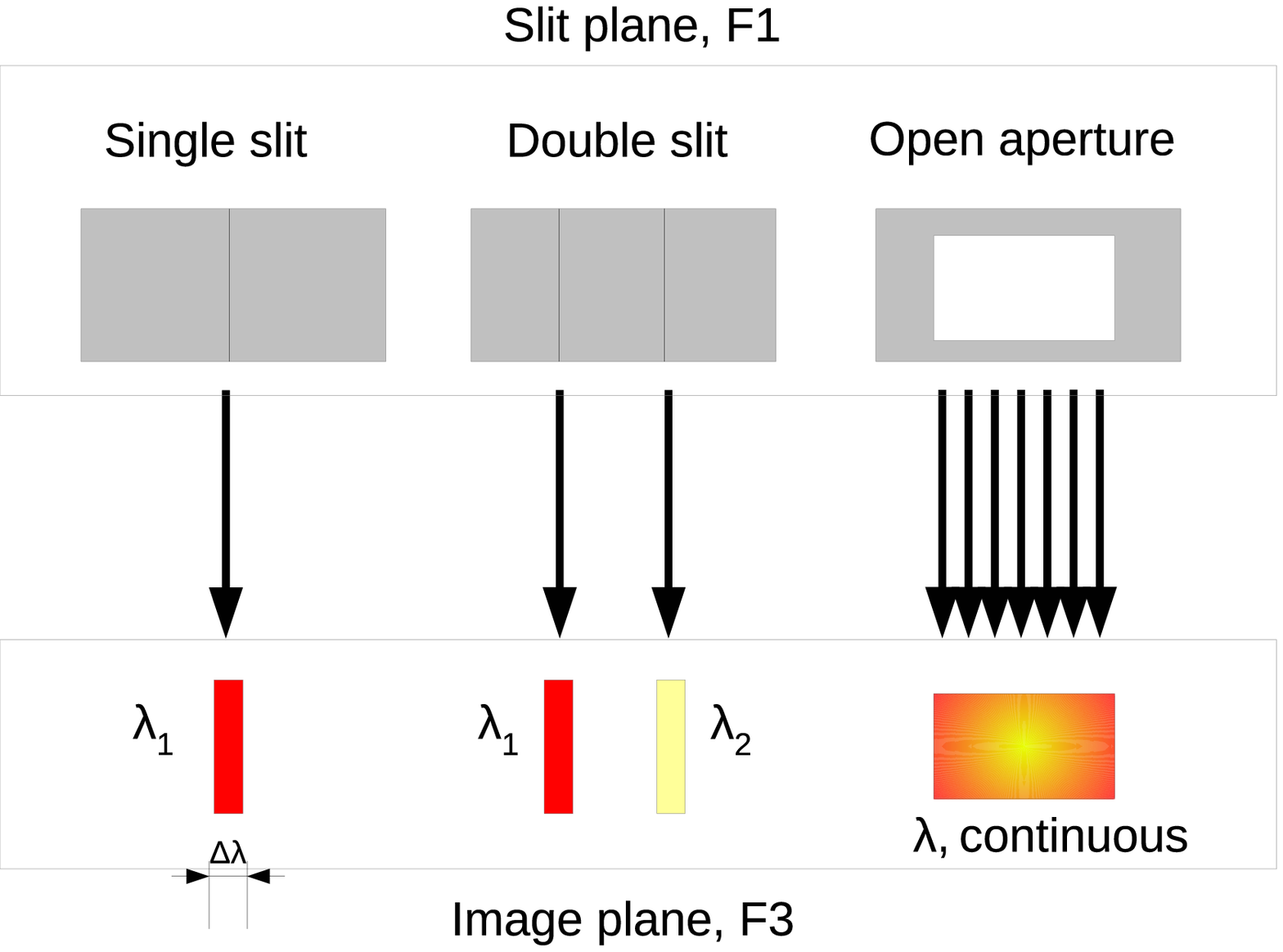}}}
\caption{Schematic concept of a SDP. Left: basic SDP concept. The light from the slit plane (focal plane F1) creates a regular dispersed spectrum at the focal plane F2 on the first pass through collimator, grating and camera mirror (or lens). The spectra from different locations in $x$ in F1 overlap, but are displaced in wavelength in F2. The wavelength is constant in $y$. The spectral slit mask only transmits a small wavelength range $\Delta\lambda$. The second pass in the reverse order of optical elements removes the spectral dispersion again (``subtraction'') and yields a narrow-band image of the slit plane F1 at the final focal plane F3. Right: wavelength variation in F3. Each spatial location in $x$ is imaged at a slightly different wavelength in the final focal plane F3. Replacing the slit unit by an open aperture, a continuous, monotonic variation of wavelength across the FOV results (see Figure \ref{sdpworks}). 
} \label{sdp_concept}
\end{figure}

A generic SDP layout is shown in the left panel of Figure \ref{sdp_concept}. White light from the entrance plane of a spectrograph (focal plane F1), where commonly a spatial slit mask is located, passes through a collimator onto a dispersing grating. The dispersed light is re-imaged onto a focal plane F2 by a camera lens or mirror. In the regular spectrograph operation, a camera in F2 would record a spectrum from the light that passed through the entrance slit. For the SDP, the camera is replaced by a spectral slit mask that cuts out a small wavelength range $\Delta\lambda$, whose extent is set by the dispersion and the physical slit width, from the spectrum. The light passing the spectral slit is reflected back towards the grating through either the prior camera lens -- now acting as collimator -- or a second separate collimating optics. As the grating is now passed a second time in the reverse order to the incoming beam, it removes the wavelength dispersion, which constitutes the so-called ``subtractive'' double pass. The return beam gets imaged by the previous collimator onto a focal plane F3 at the location of the entrance slit plane F1. If the back-reflection happens with a small tilt angle, the image in F3 is displaced from F1 and thus can be recorded without vignetting the initial incoming beam. 

The main difference of SDP and FPI-based systems is that the former do not sample the same wavelength across the FOV, even if the wavelength band-pass $\Delta\lambda$ is identical. The latter show some variation in the transmitted wavelength if they are used in a collimated beam, but its magnitude is much smaller than in SDP systems. For a conventional single-slit spectrograph, the spectrum in the focal plane F2 can be imagined as a series of images of the entrance slit at different wavelengths. If one now imagines the slit plane F1 to consist of a number of consecutive individual ``slits'' next to each other in $x$, each such ``slit'' in F1 produces a spectrum in F2 that is laterally displaced relative to the others. As only one spectral slit mask is used, a slightly different wavelength is transmitted for each ``slit'' in $x$, while the wavelength along the $y$ direction is constant for a given $x$. Replacing the entrance slit plane with an open aperture then leads to a smooth and monotonic variation of the wavelength in F3 in the $x$-direction (right half of Figure \ref{sdp_concept} and Figure \ref{sdpworks} below). 

For a given orientation of the grating or a given location of the spectral slit mask, only one narrow-band wavelength is sampled on each column of any camera at F3. To observe a spectrum over the full extent of F3, the spectral range or line of interest has to be moved once across F3. The two possible options are to move the spectral slit mask laterally or to change the grating angle, whichever is technically feasible. This leads to a sequential observation with multiple steps for acquiring specific wavelengths like for FPI systems with the main difference that the sampled wavelengths strongly vary across the FOV in each image for an SDP.

\section{Observational Data}
\begin{figure}
\centerline{\resizebox{13.cm}{!}{\includegraphics{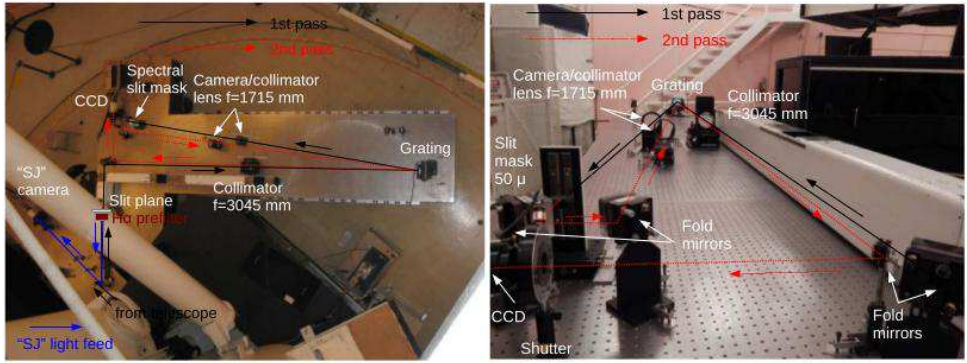}}}
\caption{Pictures of the SDP setup at the DST. Left (right) panel: top (front) view. The beam from the telescope encounters an H$\alpha$ interference filter at the regular slit plane. The reflected beam is used to feed a ``slit-jaw'' imager (blue lines in the left panel). The H$\alpha$ light transmitted through the pre-filter (black lines and arrows) passes through collimator, grating and camera lens and creates a dispersed spectrum on the spectral slit mask (see Figure \ref{setup_details}). The light passing the spectral slit is folded back onto the grating through a second collimator lens (red lines and arrows). The second pass is reflected on the grating and the standard collimator now acts as a camera lens instead. The beam of the second pass is slightly laterally displaced from that of the first pass. A small pick-up mirror close to the first fold mirror of the spectrograph reflects the second pass to the side onto the camera.}\label{setup_pics} 
\end{figure}  
\subsection{Subtractive Double Pass}\label{sdp_realize}
\subsubsection{Setup}
Following the generic layout of an SDP in Figure \ref{sdp_concept}, we converted the HSG at the DST into an imaging spectrometer for an observing campaign in 8--17 July 2015. Figure \ref{setup_pics} shows the HSG SDP setup. Only optical components that were readily available at the DST were used (see Figure \ref{setup_details}). The HSG was initially set up for regular observations in H$\alpha$ using a spatial slit mask. We then folded the light back to the entrance aperture, but because of the diameter of the HSG camera lenses, the collimator and the grating it was impossible to use the camera lens of the first pass as collimator for the second pass. We thus had to use a second lens as collimator for this purpose. Additionally, we had to displace the second pass slightly from the central optical axis to separate it from the first pass close to the entrance slit plane. That led to some vignetting of the pupil image of the second pass on the grating (see Appendix \ref{sdp_setup_details}). With the optics for the return beam in place and aligned, we added the spectral slit mask at F2. 
\begin{figure}
\centerline{\resizebox{13.cm}{!}{\includegraphics{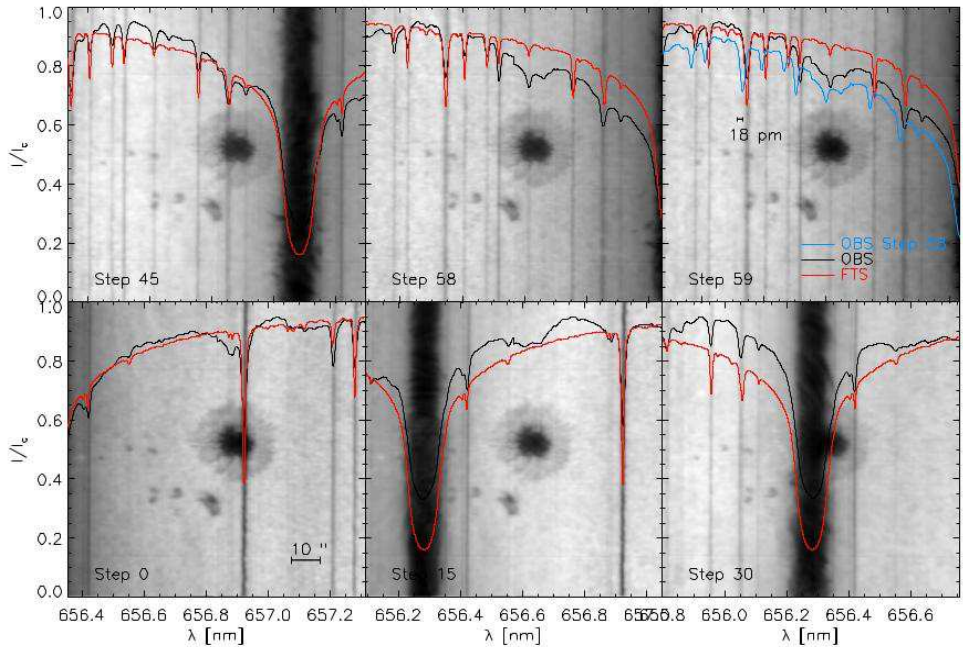}}}
\caption{Execution of a SDP spectral scan. The H$\alpha$ line core was moved once across the CCD from left to right in 60 steps. Bottom row, left to right: recorded images for steps 0, 15 and 30. Top row, left to right: the same for steps 45, 58 and 59. Red line: FTS atlas. Black line: observed profile averaged along the $y$-axis.The intensity of the observed profiles is slightly reduced each time on the columns where the sunspot was located because of the averaging in $y$. Blue line for step 59: average profile of step 58.  The finite minimal step width of the grating led to a spectral sampling of about 18\,pm.}\label{sdpworks}
\end{figure} 

The HSG setup required an order-sorting interference filter centred on H$\alpha$. As the location of this filter in the beam, either close to the camera or upfront, has no impact on its performance, we placed it at the location of the usual entrance slit plane. We then fed a sort of ``slit-jaw'' (SJ) imager with the back-reflection from the front side of the pre-filter (see Appendix \ref{sjimager}). We note that we could have reflected this beam also to another post-focus instrument. This option is, however, only possible as long as a only single spectral line is observed with the SDP. We did not use the images from the SJ imager in the current study, but the example images in Figure \ref{sj} show that the spatial resolution and image quality at the entrance plane of the spectrograph was very good.

\subsubsection{Technical Characteristics and Operation}
The plate scale on the HSG entrance (F1) is 7.5$^{\prime\prime}$\,mm$^{-1}$. Since the first/second pass used camera/collimator lenses with an identical focal length of 1.7\,m, the plate scale in F3 was almost identical to that of F1. We used a 1k x 1k camera with 13-$\mu$m pixels in F3 that gave a total FOV of about 100$^{\prime\prime} \times 100^{\prime\prime}$ at a spatial sampling of about 0\farcs1. 

We used a grating with 308 grooves\,mm$^{-1}$ and a f=1.7\,m camera lens to create an 8th-order spectrum at 656\,nm at F2 with a linear dispersion of about 0.16 pm\,$\mu$m$^{-1}$. The 50-$\mu$m spectral slit mask thus transmitted a band-pass $\Delta\lambda$ of about 8\,pm. The slit being a physical mask, this is a ``hard'' band-pass, not a measurement of full width at half maximum. In principle, we could have used the standard HSG slit aperture that has an adjustable slit width because the HSG slit unit can be taken off, but it turned out to be too bulky to be mounted easily at F2. 

\begin{figure}
\centerline{\begin{minipage}{8cm}\resizebox{8cm}{!}{\includegraphics{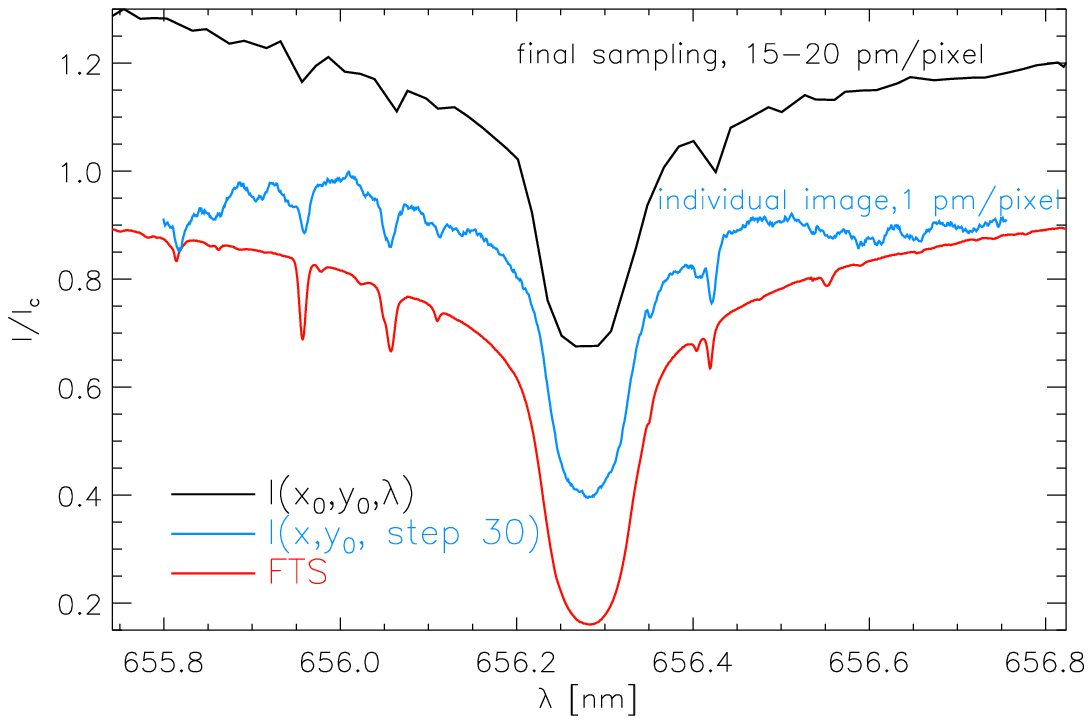}}\end{minipage}\hspace*{1.cm}\begin{minipage}{5cm}\resizebox{5cm}{!}{\includegraphics{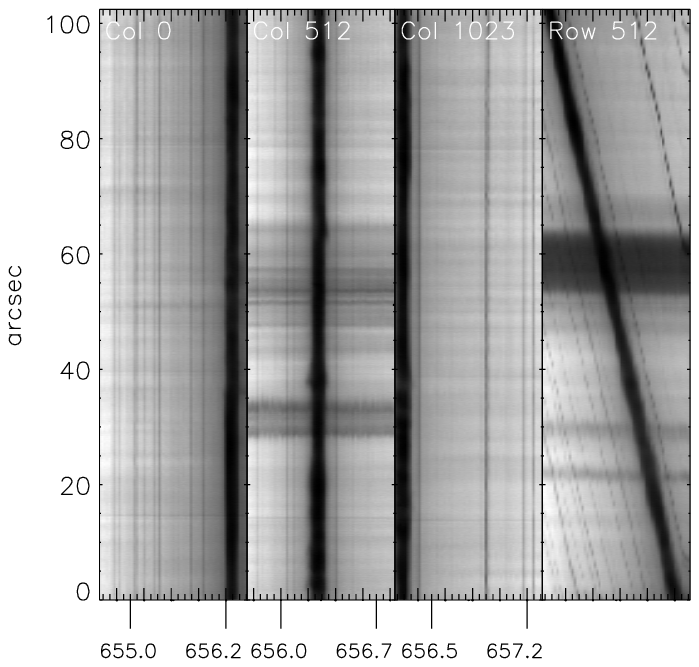}}\end{minipage}}$ $\\
\caption{SDP spectral performance. Left panel: spectra corresponding to a single spatial position $x_0, y_0$ (black line), a row of a single image at a given grating angle (blue line) and the FTS atlas spectrum (red line). The profiles have been displaced in $y$ for clarity. Right panel, left to right: spectra on spatial cuts across the FOV along the $y$-axis on column 0 (left border of FOV), 512 (middle) and 1023 (right border of FOV), and spectra on a cut across the FOV along the $x$-axis on row 512 (middle of FOV).}\label{specshape}
\end{figure}

Without a motorized spectral slit mask, we had to rotate the grating for spectral scans. The minimal reliable step size of the HSG grating motor turned out to about 0.02 degrees. Smaller steps could be requested, but were often not executed at all. Figure \ref{sdpworks} shows the motion of the H$\alpha$ line across the CCD at F3 in 60 steps of 0.02\,deg step width in grating angle, which was one of the typical settings that we ran. The grating motion translated into a final spectral sampling of about 15--20\,pm (Figure \ref{sdpworks}) while individual images contain a spectrum through an 8\,pm-band-pass on a 1 pm\,pixel$^{-1}$ sampling. The spectral sampling and the effective spectral resolution in individual images compares fairly well with the Fourier Transform Spectrometer atlas \citep[FTS;][]{kurucz+etal1984}, but the line profiles get under-sampled in the spectral scanning (left panel of Figure \ref{specshape}). The full wavelength range covered in each spectral scan was about 1\,nm.  

The variation of wavelength across the FOV in $x$ in an SDP leads to peculiar data sets. Every column in the FOV at F3 has a different wavelength scale with a nearly identical dispersion, but a varying zero point (right panel of Figure \ref{specshape}). The spectral line of interest is only covered on all columns if the spectral scanning moved it across the full CCD. This drawback can also be converted to an advantage because one can achieve a faster cadence with a restriction of the FOV covered (see Appendix \ref{highcadence}). The rightmost sub-panel of the right panel of Figure \ref{specshape} shows the spectra along a row of the CCD at F3. It visualizes the most compact description of the data produced by an SDP: ``An SDP is an imaging spectrometer that makes an inclined cut through a rectangular $x$-$y$-$\lambda$ data cube''\footnote{Phrase attributed to A.~Lopez Ariste.}.

The spatial resolution in individual images (Figure \ref{sdpworks}) is sufficient to identify umbral dots and penumbral filaments. A determination of the Fourier power spectrum in the spatial domain in images with a fixed wavelength across the FOV yielded a spatial resolution of about 0\farcs8--1$^{\prime\prime}$ (see Appendix \ref{spatres}). The degradation from diffraction-limited performance presumably comes from some astigmatism of unknown origin that is only present in the SDP setup because it is missing in the ``SJ'' images (Figures \ref{sdp_resolution} and \ref{sj}). 

Without any dedicated control software for the SDP setup, the observations had to be done with a manual synchronization between the grating position and the exposures. We ran the camera at F3 at a fixed frame rate with initially a 2.5\,s wait time between exposures of 100--400\,ms, and stepped the grating by one step of 0.02\,deg between two exposures. The ``synchronization'' to the exposures was ensured acoustically by listening to the sound of the mechanical shutter, which was complicated by the fact that the SJ camera used an identical device. After turning the SDP shutter control box around, we were able to use its LED instead, which allowed us to reduce the wait time to 1.5\,s. To increase the cadence further, the spectrum was repeatedly scanned back and forth across the CCD for ten repetitions instead of starting off from the same edge of the FOV each time. This led to a cadence for a single spectral scan of 60--75 steps of about 90--150\,s. With an actual control software and a faster camera, the temporal performance of an SDP could be matched to that of an FPI-based system. The only mechanical motion is the rotation of the grating or the motion of the spectral slit mask, while the needs for the control of exposures are identical to those in an FPI instrument.

\subsubsection{Example SDP Observations}
\begin{figure}
\centerline{\resizebox{13.cm}{!}{\includegraphics{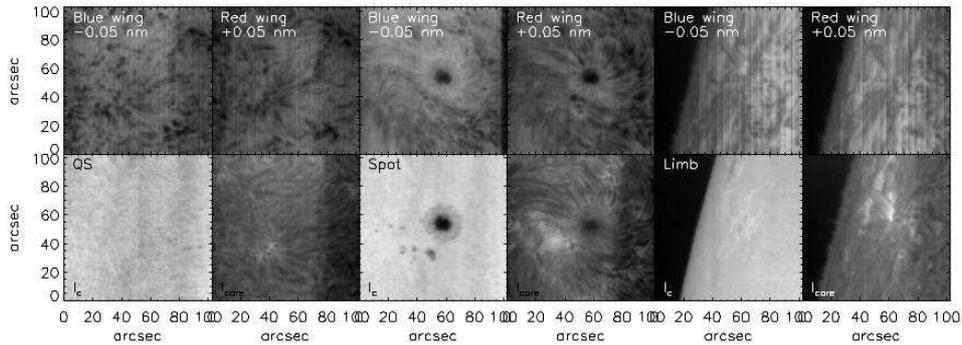}}}
\caption{Example SDP data. Left to right: quiet Sun at disc center recorded on 13 July 2015 UT 15:37, sunspot recorded on 13 July 2015 UT 15:55, and solar limb recorded on 15 July 2015 UT 15:35. For each observation, the panels show in clockwise order, starting left bottom, the continuum intensity, the intensity in the blue and red wing, and the line-core intensity. } \label{dataexam}
\end{figure}

We obtained data with the SDP setup at the DST from 2015 July 12--16. Figure \ref{dataexam} shows three examples of SDP observations on different solar targets (quiet Sun, sunspot, limb). The maps in the blue and red wing show a pattern of vertical stripes that is absent in the line-core and continuum intensity map. The stripes are caused by the combination of sequential wavelength scanning, a coarse and non-equidistant spectral sampling, the attempt to retrieve a specific wavelength and the slope of the spectrum at that wavelength. The spectrum at the line core and continuum wavelengths is basically flat, so the exact wavelength is uncritical, whereas in the line wing the intensity gradient causes jumps if the wavelength varies only a little. The effect can be mitigated by interpolating the spectra to a denser equidistant wavelength sampling before (see Section \ref{sdp_ana} below). Otherwise, the spatial resolution of the data is fully satisfactorily, as given, \textit{e.g.}, by the sharpness of the limb, the lateral extent of filaments in the QS and sunspot data or the numerous bright grains around the sunspot just at the end of the penumbra in its blue-wing image. We note that the limitation to a $100^{\prime\prime}\times 100^{\prime\prime}$ FOV was primarily due to the lack of a larger CCD. The DST light feed and the HSG aperture allows one to observe up to a (3 arcmin)$^2$ FOV.
\subsubsection{SDP Data in Current Study}
For the current study, we concentrate on a set of sunspot observations taken on 2015 July 13 from UT 15:55--16:26 corresponding to the middle panels of Figure \ref{dataexam}. We obtained in total 10 spectral scans of 60 grating steps each with a cadence of about 3 min. The target was NOAA 12384 close to the central meridian at about $(x,y) = (0^{\prime\prime},- 380^{\prime\prime})$. The corresponding active region (AR) was already in the decaying stage at that time, with a leading but no follower sunspot and a network of opposite polarities spread out over a few degrees in longitude and latitude. There was some flaring activity towards the east and south of the DST FOV outside our observing time slot. A pronounced filament related to the AR was present but also outside of the FOV at the DST (see Figure \ref{full_disk}).
\subsection{Synoptic Data}
For comparison to the SDP and to increase the information on superpenumbral filaments, we selected a set of complementary full-disc observations. We retrieved full-disc images in H$\alpha$ from the \textit{Global Oscillation Network Group} \citep[GONG;][]{harvey+etal1996}, photospheric line-of-sight (LOS) magnetograms and LOS velocities from the \textit{Helioseismic and Magnetic Imager} \citep[HMI;][]{scherrer+etal2012}, images at 304\,nm from the \textit{Atmospheric Imaging Assembly} \citep[AIA;][]{lemen+etal2012} and the Stokes I and V spectra of \ion{Ca}{ii} IR at 854.2\,nm from the \textit{Synoptic Optical Long-term Investigations of the Sun} \citep[SOLIS;][]{keller+etal2003}. Apart from SOLIS, we selected the images closest in time to each SDP scan whereas the former only provides one daily measurement of spectra. The spatial sampling of the SOLIS and GONG data is about 1$^{\prime\prime}$, while the HMI and AIA data have about 0\farcs5 per pixel. The temporal cadence varied from a few to a few ten minutes. 
\section{Data Analysis}
\subsection{Inversion of SOLIS \ion{Ca}{ii} IR Spectra}
We used the \textit{CAlcium Inversion using a Spectral ARchive} code \citep[CAISAR;][]{beck+etal2013a,beck+etal2013b,beck+etal2015} to derive temperature stratifications from the photosphere to the chromosphere from the SOLIS full-disc spectra of \ion{Ca}{ii} IR at 854.2\,nm. These data were taken around UT 18:05 on 13 July 2015, about two hours after the DST observations. The inversion provides the temperature stratifications as a function of the logarithm of optical depth, log $\tau$, at 500\,nm. The CAISAR code has originally been developed for observations at disc center, so the inversion results become a bit more coarse towards the limb (Figure \ref{full_disk}) because the archive lacks temperature stratifications that correspond to an inclined LOS. The target region that we focus on here was, however, comparably close to disc center. Unfortunately, the gap between the two SOLIS CCDs passed through the region as well. That causes a stripe of reduced to zero temperature close to the sunspot which shows up prominently in all graphs based on the inversion results.  
\subsection{Reduction and Analysis of SDP Spectra}\label{sdp_ana}
The SDP data can to some extent be treated like regular spectrograph observations. We took standard flat field data with the grating set such as to have a wavelength region without the H$\alpha$ line and with mainly continuum wavelengths in the image. Together with a dark current measurement, this provided a gain correction for the wavelength-independent inhomogeneities across the CCD like for regular slit spectrographs. 

The SDP images have a different wavelength scale across the image on every step (Figure \ref{sdpworks}). We used the spectrum from the FTS atlas to determine the wavelength scale in each image. After determining the dispersion in the SDP images, we correlated the average profile of each SDP image in $y$ with the FTS spectrum. The correlation alone was not fully reliable because of instrumental or true (H$\alpha$ line core, location of sunspot) intensity gradients in the spectrum. We thus had to use an interactive, semi-automatic method with the option to manually provide zero points for the wavelength scale of the SDP to properly match the FTS spectrum. Given that the step width of the grating was halfway constant, even the fully manual mode for the determination of the wavelength scale was manageable. 

For the determination of characteristic properties of H$\alpha$ in each spectrum, we spectrally re-sampled the SDP data of 60 non-equidistant wavelength points to an equidistant grid of 100 points. We then determined the continuum and line-core intensity and the line-core velocity of the H$\alpha$ line. For the continuum intensity, we used a wavelength point in the blue or red line wing depending on the number of the column on the CCD as the observed wavelength range varies (\textit{e.g.}, Figure \ref{specshape}). The line-core intensity was derived from the point of minimal intensity in the re-sampled spectra, while for the line-core velocity we used a Fourier-based determination of this location with sub-pixel precision. We fitted a Gaussian to the surroundings of the H$\alpha$ line core for an estimate of the line width. This fit could not be done for columns at the very edge of the FOV in some cases, \textit{e.g.}, at the left border of the line-width maps in Figure \ref{ts1} below, when the H$\alpha$ line core had not been moved completely off the CCD far enough. 

\subsection{Data Alignment}
For all of the full-disc data with multiple images, we first aligned each series internally using a cut-out of a few hundred arcseconds extent around NOAA 12384. We then aligned them to the SDP data based on whatever quantity from the H$\alpha$ spectra worked best, usually either the continuum or the line-core intensity. Full-disc data was aligned to each other on the basis of the solar radius and the disc centre location.
\section{Results}\label{secresults}
\begin{figure}
\centerline{\resizebox{10cm}{!}{\includegraphics{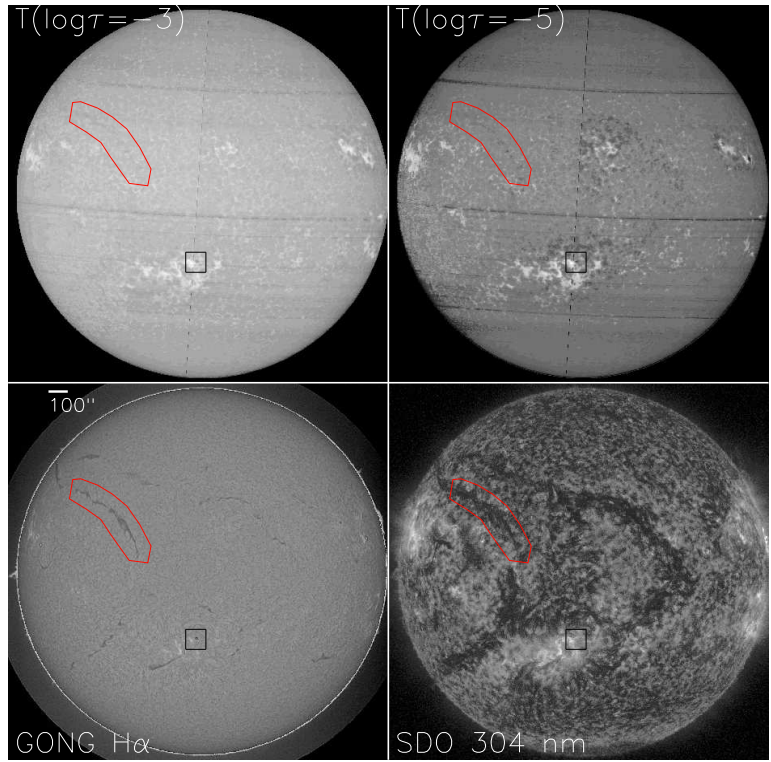}}}
\caption{Full-disc overview images. Bottom row: GONG H$\alpha$ (left) and AIA 304\,nm images (right). Top row: temperature at log $\tau = -3$ and $-5$ from the inversion of the SOLIS data. The red polygon outlines the largest filament in the GONG H$\alpha$ image. The black square indicates the FOV of the SDP observations. The nearly vertical black dashed line near the meridian in the SOLIS images is an instrumental artifact in the SOLIS data.}\label{full_disk}
\end{figure}   

\begin{figure}
\centerline{\resizebox{13cm}{!}{\includegraphics{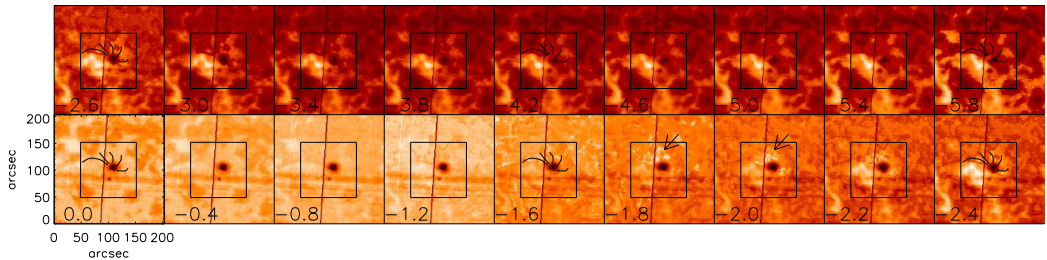}}}$ $\\
\caption{Temperature maps of the active region from the inversion of the SOLIS data. The labels in the lower left corner of each sub-panel indicate the log $\tau$ level from 0 to $-5.8$. The black square marks the FOV of the SDP observations. The curved lines in every fifth sub-panel show a few of the filament tracks of Figure \ref{burst2exam} below. The black arrows in the sub-panels of log $\tau = -1.8$ and $-2.0$ indicate a region close to the sunspot with an increased temperature in the upper photosphere.} \label{solis_invert}
\end{figure}

\subsection{Comparison of Full-disc Data}
To investigate the suitability of the GONG H$\alpha$, SOLIS \ion{Ca}{ii} IR and AIA 304\,nm data for determining the properties of superpenumbral filaments, we first attempted to verify that such structures can be seen in them. From the full-disc images in Figure \ref{full_disk} it became already clear that this only works to some extent. The GONG H$\alpha$ data is lacking weaker filaments such as those in the superpenumbra, while in the SOLIS \ion{Ca}{ii} IR data even the strongest filaments in the GONG H$\alpha$ data only show up faintly (red polygon in Figure \ref{full_disk}). The spatial patterns in AIA 304\,nm match to some extent better to the SOLIS \ion{Ca}{ii} IR data than the GONG H$\alpha$ line-core images. Some part of the increased temperature in the chromosphere and the increased emission in the AIA 304\,nm data close to and within the SDP FOV is the residual of flaring activity in that region. From the full-disc images, we had to conclude that those channels will not be able to provide much information on the small-scale, faint filaments in the superpenumbra but we still kept some of them in the following for completeness. We note that a more detailed comparison of the different channels with the full-disc temperature stratifications might be interesting, but this is beyond the scope of the current study \citep[see also][]{pietarila+harvey2013}.

\subsection{Temperature Maps from Inversion of \ion{Ca}{ii} IR Spectra}
Figure \ref{solis_invert} shows the inversion results of the \ion{Ca}{ii} IR spectra in the surroundings of the SDP observations. The temperature from log $\tau= -2$ to $-5.8$ is increased south and east of the sunspot as a result of the flaring activity, while at photospheric layers some small pores are seen in that region. The tracks of the superpenumbral filaments determined from the SDP H$\alpha$ data do not touch that region, but go off to the north and north-east. These tracks overlay a region where the temperature just outside the penumbral boundary is enhanced almost all around the sunspot in the upper photosphere at about log $\tau= -2$. No individual filaments can be discerned at layers above log $\tau= -3$ at the spatial resolution of SOLIS, but the temperature in the area of the filament tracks is lower than in most of the surroundings.
\begin{figure}
\centerline{\resizebox{13cm}{!}{\includegraphics{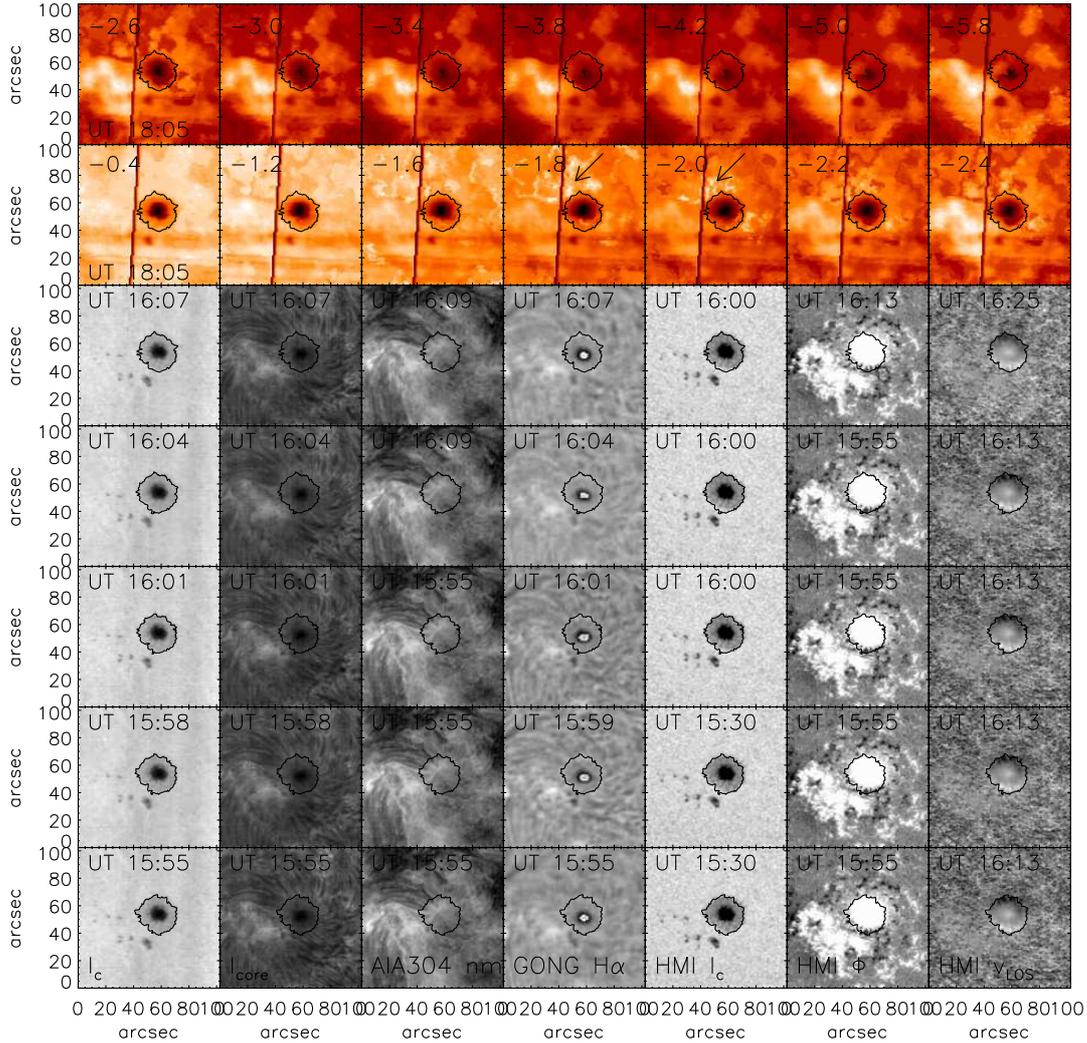}}}$ $\\
\caption{Aligned data for the SDP spectral scans 1--5 taken on 13 July 2015 from UT 15:55--16:07. Left to right: continuum intensity and line-core intensity from the SDP H$\alpha$ spectra, AIA 304\,nm cut-out, GONG H$\alpha$ cut-out, continuum intensity, LOS magnetic flux and photospheric LOS velocity from HMI. The contour line indicates the extension of the penumbra in the SDP continuum intensity map. Time increases from bottom to top. The two panels in the top row show the temperature at log$\tau = -1$ and $-4$ from the inversion of the SOLIS data.}\label{sdp_allscans}
\end{figure}

\begin{figure}
\centerline{\resizebox{13cm}{!}{\includegraphics{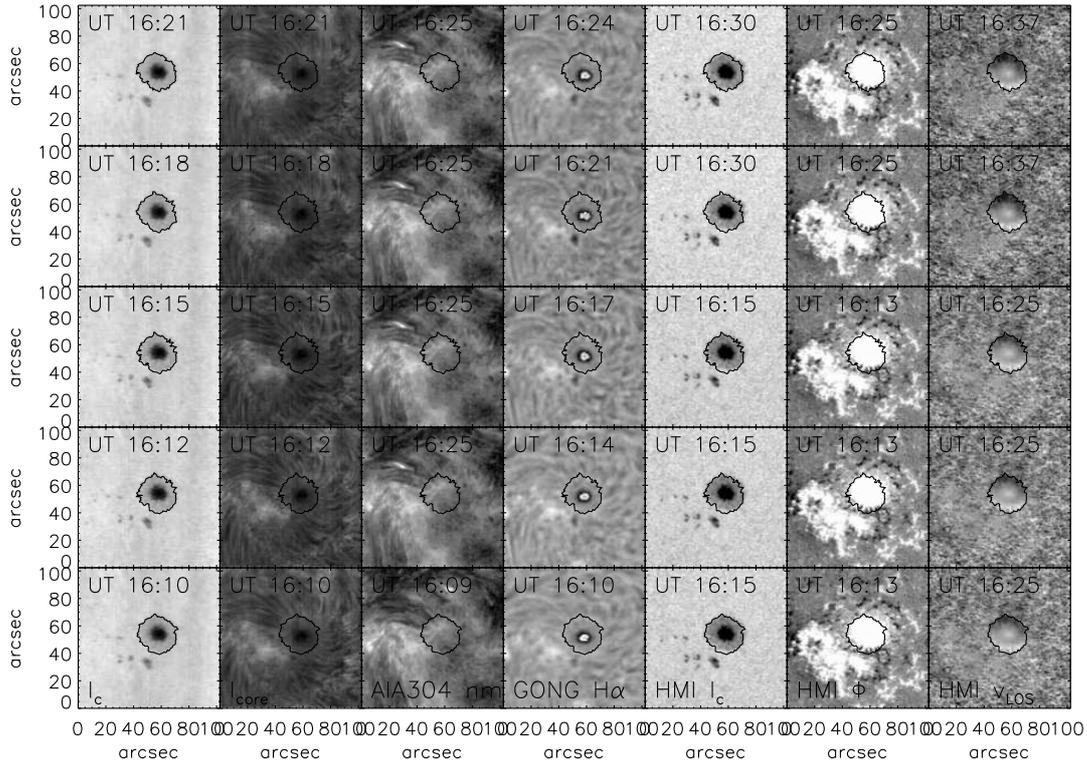}}}$ $\\
\caption{Same as Figure \ref{sdp_allscans} for the SDP spectral scans 6--10 taken on 13 July 2015 from UT 16:10--16:21.} \label{sdp_allscans1}
\end{figure}   

\begin{figure}
\centerline{\resizebox{13cm}{!}{\includegraphics{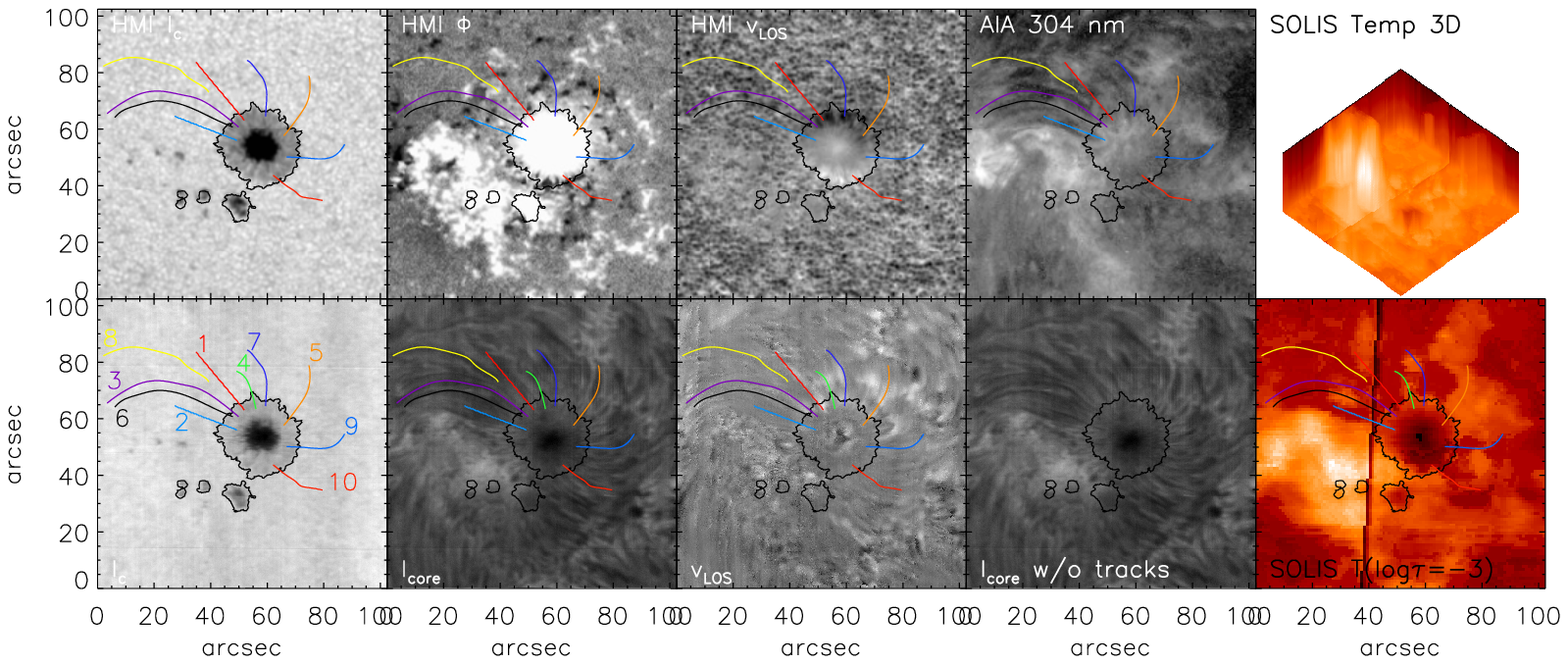}}}$ $\\
\caption{Aligned data for the third spectral scan of the SDP at UT 16:01 on 13 July 2015. Bottom row, left to right: continuum intensity, line-core intensity, line-core velocity,  line-core intensity without filament tracks (all from the SDP H$\alpha$ spectra), and temperature at log$\tau = -3$ from the inversion of the SOLIS data. Top row, left to right: continuum intensity, LOS magnetic flux, photospheric LOS velocity (all from HMI), AIA 304\,nm cut-out, and 3D rendering of the temperature cube from the inversion of the SOLIS data. In the latter, the upper (left) end of the FOV is towards the right (left) upper corner.}\label{burst2exam}
\end{figure}   
\subsection{Aligned SDP, AIA 304\,nm, GONG H$\alpha$ and HMI Data}
Figures \ref{sdp_allscans} and \ref{sdp_allscans1} show the SDP data together with aligned sub-fields of AIA 304\,nm, GONG H$\alpha$, and the photospheric continuum intensity, LOS magnetograms and LOS velocities from HMI. The AIA 304\,nm image exhibits the same superpenumbral filaments as the SDP H$\alpha$ line-core image, but only for long H$\alpha$ filaments. All short H$\alpha$ filaments towards the north and west are missing in AIA 304\,nm, the same as for GONG H$\alpha$ for almost all of the filaments. Short H$\alpha$ filaments thus seem to lack sufficient opacity to show up in the AIA 304\,nm and GONG H$\alpha$ data. The filaments are seen to extend from the positive-polarity sunspot into regions of opposite polarity flux. The region of increased temperature at log\,$\tau = -4$ in the \ion{Ca}{ii} IR inversion south-east of the sunspot shows up bright as well in all H$\alpha$ line-core and the 304\,nm images. It coincides well with strong signal in the HMI magnetograms.

A magnification of the FOV is shown in Figure \ref{burst2exam} for the third SDP scan at UT 16:01. We used this scan to trace in total ten superpenumbral filaments to determine characteristic properties along their path. The match of increased temperature and increased intensity in H$\alpha$ and AIA 304\,nm south and east of the sunspot is even more obvious there. The 3D rendering of the inversion temperature in the upper rightmost panel of Figure \ref{burst2exam} shows that this temperature increase persists throughout the chromosphere at all layers of optical depth. The superpenumbral filaments in the H$\alpha$ line-core image are mapped in the H$\alpha$ LOS velocity, but show a smoother lateral variation in it. The inner foot-points of filament tracks predominantly show down-flows as an indication of the inverse Evershed flow.

\begin{figure}
\centerline{\resizebox{15cm}{!}{\includegraphics{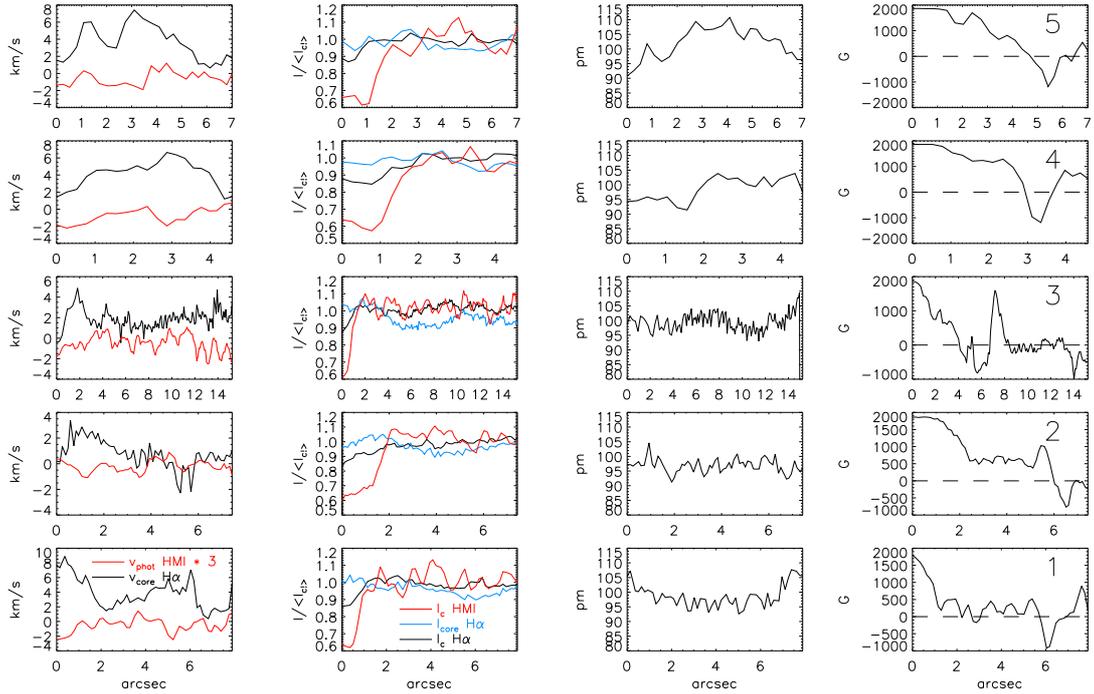}}}$ $\\
\caption{Parameters along the filament tracks marked in Figure \ref{burst2exam}. Left to right: LOS velocity of H$\alpha$ (black) and HMI multiplied by 3 (red), continuum (black) and H$\alpha$ line-core intensity (blue) from the SDP observations and HMI continuum intensity (red), line width, and LOS magnetic flux. Bottom to top: filament tracks 1--10. All tracks start at the point closest to the umbra of the sunspot. }\label{cuts1}
\end{figure}    

\subsection{Properties of Superpenumbral Filaments}
We extracted the photospheric and chromospheric LOS velocities, continuum and line-core intensities, line width and LOS magnetic flux along the ten filament tracks marked in Figure \ref{burst2exam}. The chromospheric velocities (leftmost column of Figure \ref{cuts1}) show preferentially down-flows at the inner end of the filament and along its length. The H$\alpha$ line-core intensity is naturally reduced along their length, as the filaments are defined as dark structures. The line width is generally increased along the filaments, but shows little small-scale variation. The most prominent feature is seen in the LOS magnetic flux where nearly all filament tracks end at or close to patches of magnetic fields with a polarity opposite to that of the sunspot. This supports the interpretation that H$\alpha$ filaments trace chromospheric magnetic field lines that connect opposite polarities. 

\begin{figure}
\centerline{\resizebox{8cm}{!}{\includegraphics{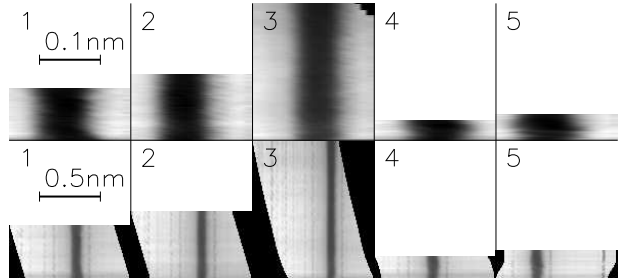}}}$ $\\
\caption{Spectra along the filament tracks marked in Figure \ref{burst2exam}. For each filament, the full spectral range of about 1\,nm is shown in the lower and a magnification of the H$\alpha$ line core in the upper row.}\label{cuts2}
\end{figure}   

The SDP H$\alpha$ spectra along the filament tracks are shown in Figure \ref{cuts2}. The tracks crossed different columns in the FOV so the spectral range sampled varies along some of the tracks. As we had, however, scanned the H$\alpha$ line completely across the CCD, the SDP data look exactly like regular spectrograph data with the only difference of covering different ranges in the blue and red wing of H$\alpha$ depending on the column the spectra were taken from. The variation in the line properties and shape along the filaments only shows up when the full spectral range of about 1\,nm is reduced to only show the H$\alpha$ line core.
\begin{figure}
\centerline{\resizebox{8cm}{!}{\includegraphics{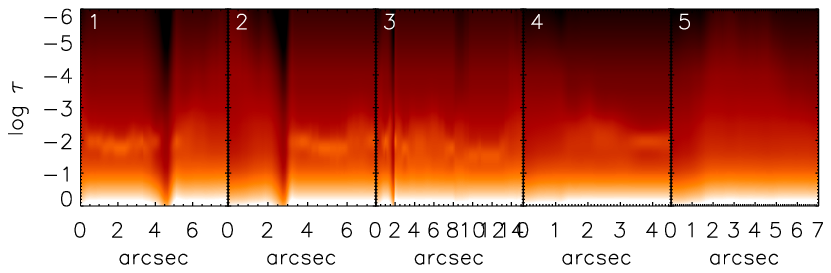}}\hspace*{.5cm}\resizebox{1.5cm}{!}{\includegraphics{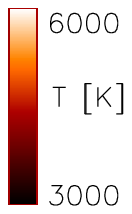}}}$ $\\
\caption{Temperature stratifications along the filament tracks marked in Figure \ref{burst2exam}. All tracks have been re-sized to an identical length for the display. The point closest to the penumbra is at the left in each sub-panel.}\label{cuts3}
\end{figure}   

The temperature stratifications along the filament tracks (Figure \ref{cuts3}) are another quantity that shows a prominent feature that seems to be characteristic for the dark H$\alpha$ filaments. As discussed above, the temperature at about log $\tau = -2$ is enhanced in the close surroundings of the sunspot (Figure \ref{solis_invert}). For nearly all of the filament tracks, this temperature increase around log $\tau = -2$ is clearly seen, often along their full length. Given the spatial resolution of the SOLIS full-disc data and the 2-hr time-lag between the SDP and SOLIS observations, we prefer, however, to verify this pattern in quasi-simultaneous spectra of \ion{Ca}{ii} IR and H$\alpha$ of higher spatial resolution to ensure it is significant.
\begin{figure}
\centerline{\resizebox{12cm}{!}{\includegraphics{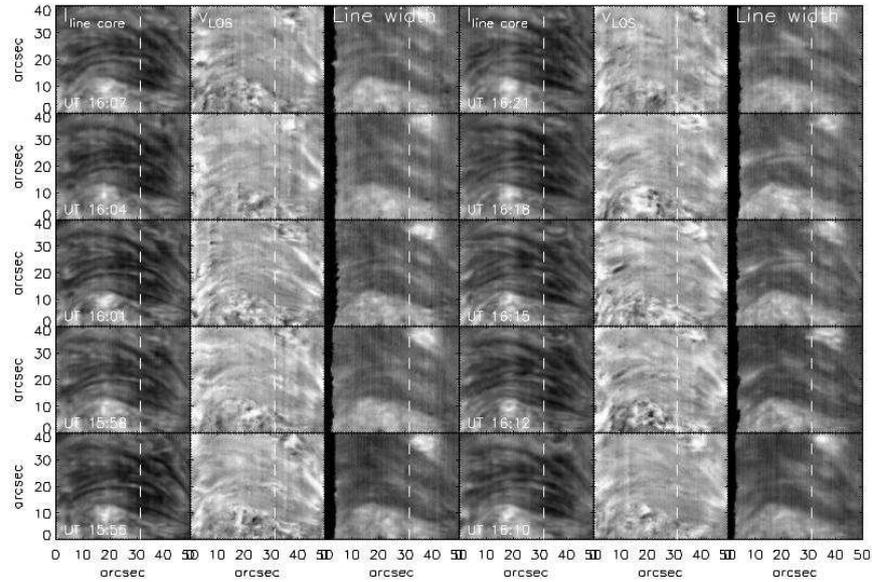}}}
\caption{Magnification of the region with the most prominent filaments. Left to right: H$\alpha$ line-core intensity, LOS velocity, and line width for the spectral scans 1--5; the same for the scans 6--10. Time increases from top to bottom in each column. The vertical dashed line indicates the location of the cuts shown in Figures \ref{ts2} and \ref{ts3}. }\label{ts1}
\end{figure}   

\begin{figure}
\centerline{\resizebox{13cm}{!}{\includegraphics{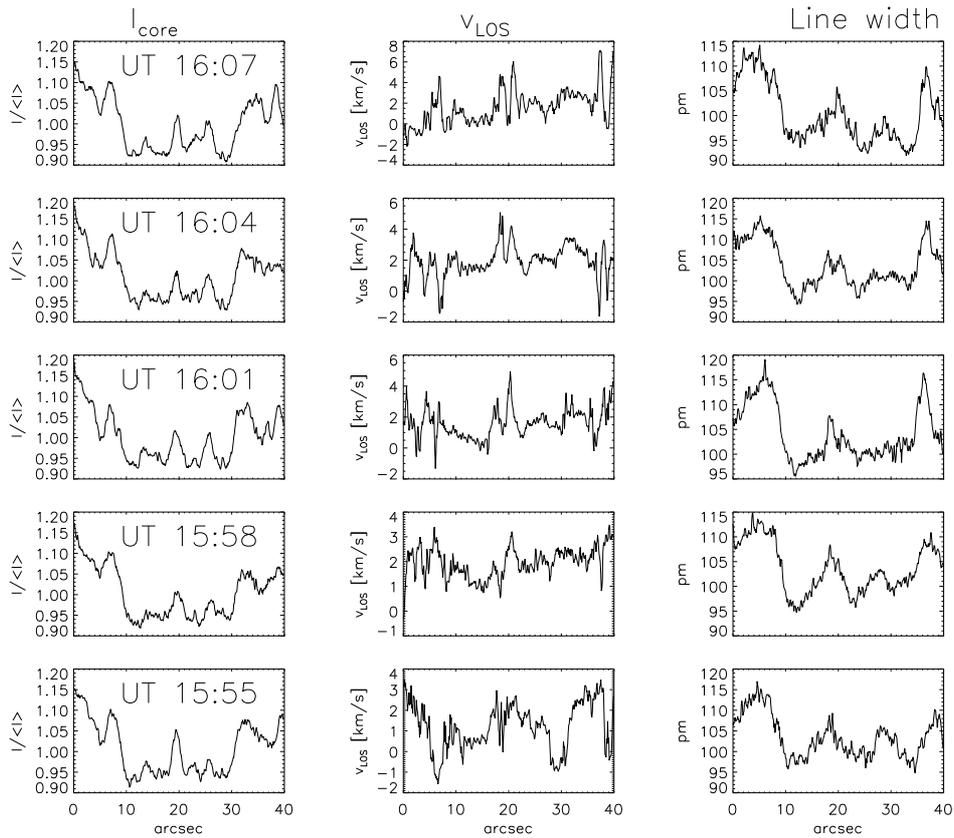}}}$ $\\
\caption{Parameters along the vertical cuts marked in Figure \ref{ts1}. Left to right: H$\alpha$ line-core intensity, LOS velocity, and line width for the spectral scans 1--5. }\label{ts2}
\end{figure}   
\subsection{Temporal Evolution and Spatial Fine-structure}
 Figure \ref{ts1} shows a close-up of the region with the darkest and longest filaments in the SDP spectra. As all displayed quantities were taken from the same set of spectra, it is ensured that they share the same spatial resolution. It is obvious that the lateral fine-structure perpendicular to the filaments in the line-core intensity is not matched in the LOS velocity or the line width. Especially in the latter, fine-structure at the sub-arcsecond scale is absent. This suggests the presence of density, and hence opacity variations seen in intensity inside a larger-scale structure with common properties instead of an ensemble of separate, independent fine threads. Most of the intensity filaments can be traced visually throughout the full sequence of about 30 min duration with only minor changes in location or shape. 

\begin{figure}
\centerline{\resizebox{13cm}{!}{\includegraphics{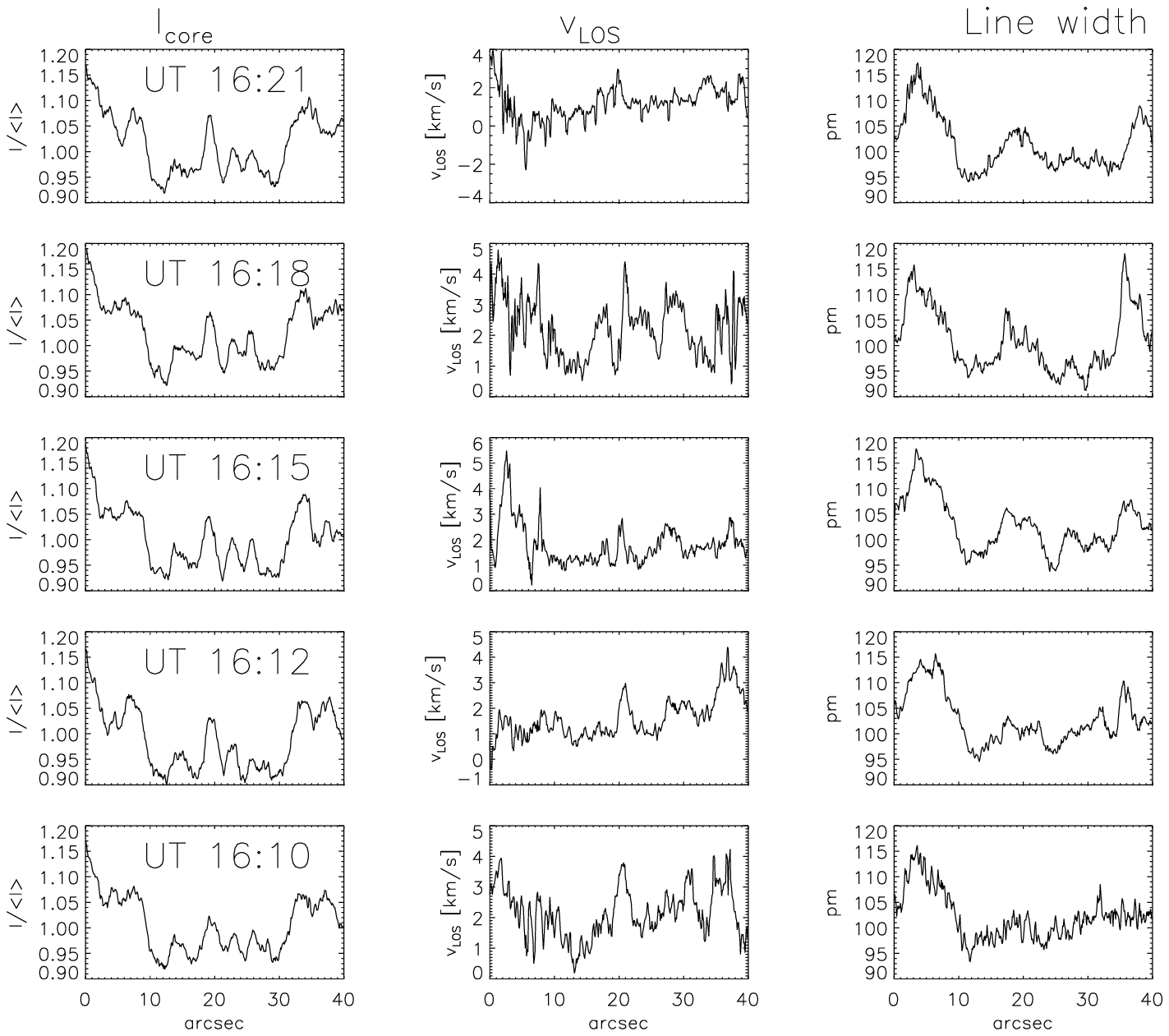}}}$ $\\
\caption{The same as Figure \ref{ts2} for the spectral scans 6--10.}\label{ts3}
\end{figure}   

The same pattern is seen in the vertical cuts across the FOV that are roughly perpendicular to the filament axes' (Figures \ref{ts2} and \ref{ts3}). Locations of increased line width extend over spatial areas of 5$^{\prime\prime}$--10$^{\prime\prime}$ in which multiple smaller-scale peaks in intensity or velocity can be found. Velocity fluctuations happen generally on shorter time-scales than variations in intensity. With the assumption that the filaments trace magnetic field lines, this points to a magnetic configuration with structures of a few arcsecond lateral extent that are stable over 30 mins, a slow variation of density over several minutes and a fast variation of flow velocities over less than three minutes in some cases.
\begin{figure}
\centerline{\resizebox{5cm}{!}{\includegraphics{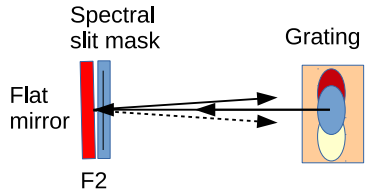}}}
\caption{Sketch of an improved SDP concept for over-sized grating and collimator. The first pass could be reflected back using a mirror at the F2 spectral focal plane with a slight upwards or downwards tilt. The second pass would be picked up above or below the incoming beam near F1. If the grating size allows one to place more than two pupil images in $y$, a multi-line setup with three (or more) lines can be realized.}\label{improved_SDP}
\end{figure}   
\section{Summary and Discussion}\label{summ}
The main purpose of our observing run was to test whether an SDP setup can be realized at the DST to provide high-resolution imaging spectroscopic observations with a conventional slit spectrograph. Apart from the issues with the astigmatism, we were able to achieve a spatial resolution below 1$^{\prime\prime}$ over a $100^{\prime\prime}\times 100^{\prime\prime}$ FOV using standard optical components that had not been designed in any way for an use in an SDP. We think that the quality of the SDP data (Figures \ref{sdpworks}, \ref{dataexam} and \ref{burst2exam}) demonstrates that SDP systems can perform as good as comparable FPI-based systems if the spatial resolution of the input image is improved by adaptive optics. 

SPD systems offer a high amount of flexibility to vary spatial, spectral or temporal sampling. They can provide a large FOV at a cadence similar to that of FPI systems, but can be sped up significantly by restricting the FOV in one axis (\textit{e.g.}, Appendix \ref{highcadence}). The resulting $x-y-\lambda$ data cubes are similar to conventional slit-spectrograph or imaging spectroscopic data apart from having to always explicitly consider the wavelength (Figures \ref{specshape} and \ref{cuts2}). SDP systems can be used at wavelengths from the near-ultraviolet into the near-infrared with the same optical components. The observed wavelength range around each spectral line of interest can be 1\,nm or more.

SDP systems can also provide a truly simultaneous multi-line capability that FPI systems are lacking and which is crucial for chromospheric studies. In case that the grating and initial collimator of a spectrograph are over-sized in the sense of being about 50\,\% larger than needed for a single pupil image, a multi-line SDP, or at least dual-channel SDP system can be easily realized by displacing the second pass(es) in the vertical instead of the horizontal direction (Figure \ref{improved_SDP}).

We note that in principle one could as well have prepared the SDP system for dual-beam polarimetry by placing a polarizing beam splitter (BS) near the last focal plane F3 and a modulator at any suited place upstream of it, \textit{e.g.}, in front of the spectrograph or just in front of the BS. This would have converted the HSG into an imaging spectropolarimeter. In that case, one would, however, have definitely needed an explicit control software for the setup to synchronize the modulation and the exposures and to do a spectral scan with the grating instead of a spatial scan of F1.
 
Some of the biggest advantages of SDP systems lie on the technical side. The necessary optical components are comparably cheap and readily available. One major point of cost saving is that only order-sorting prefilters are needed (up to 50\,nm wide), whereas FPI system need extremely narrow-band filters ($< 1$\,nm) to suppress the side-lobes in the transmission profile. These narrow filters also only work for one specific line. Operation and control of an SDP system is also much less involved than for FPI systems which implies a reduced cost in the construction and operation. Finally, a spectrograph with an added SDP-system can easily be reverted to a regular slit-spectrograph, and \textit{vice versa}, which increases the amount of possible options even further.

For the case of the properties of superpenumbral filaments that we touched on in the investigation of the SDP performance, we find no indication that dark filaments in H$\alpha$ would not trace magnetic field lines. The magnetic field configuration of the strongest filaments close to the sunspot seems to be fairly stable over at least 30 min with variations of thermodynamic properties on faster and smaller temporal and spatial scales. 

\section{Conclusions}\label{sec_concl}
We suggest that the concept of converting conventional spectrograph systems into imaging spectro(polari)meters through the use of a subtractive double pass setup should be seriously considered as an alternative for future solar telescopes larger than the 1-m class. Compared with FPI-based instruments that get increasingly difficult and expensive to built with larger telescope diameters, SDP-based instruments can be realized with a larger FOV, similar spectral and spatial sampling, and a similar cadence at much lower cost and effort. They benefit from improvements on the technical side such as larger and faster detectors in the same way as FPI systems.

\begin{acknowledgements}
This paper is dedicated to the memory of M. Bradford (NSO) and J. Staiger (KIS). The Dunn Solar Telescope at Sacramento Peak/NM was operated by the National Solar Observatory (NSO). The NSO is operated by the Association of Universities for Research in Astronomy (AURA), Inc.~under cooperative agreement with the National Science Foundation (NSF). This work utilizes GONG and SOLIS data obtained by the NSO Integrated Synoptic Program (NISP). HMI/AIA data are courtesy of NASA/SDO and the HMI/AIA science team. R.R.~acknowledges financial support by the Spanish Ministry of Economy and Competitiveness through project AYA2014-60476-P. The Center of Excellence in Space Sciences India is funded by the Ministry of Human Resource Development, Government of India.\\
$ $\\
{\bf Disclosure of Potential Conflicts of Interest} The authors declare that they have no conflicts of interest.
\end{acknowledgements}
\bibliographystyle{spr-mp-sola}
\bibliography{references_luis_mod1}
\begin{appendix}
\section{Subtractive Double Pass}
\subsection{Internal Setup}\label{sdp_setup_details}
\begin{figure}
\centerline{\resizebox{7.5cm}{!}{\includegraphics{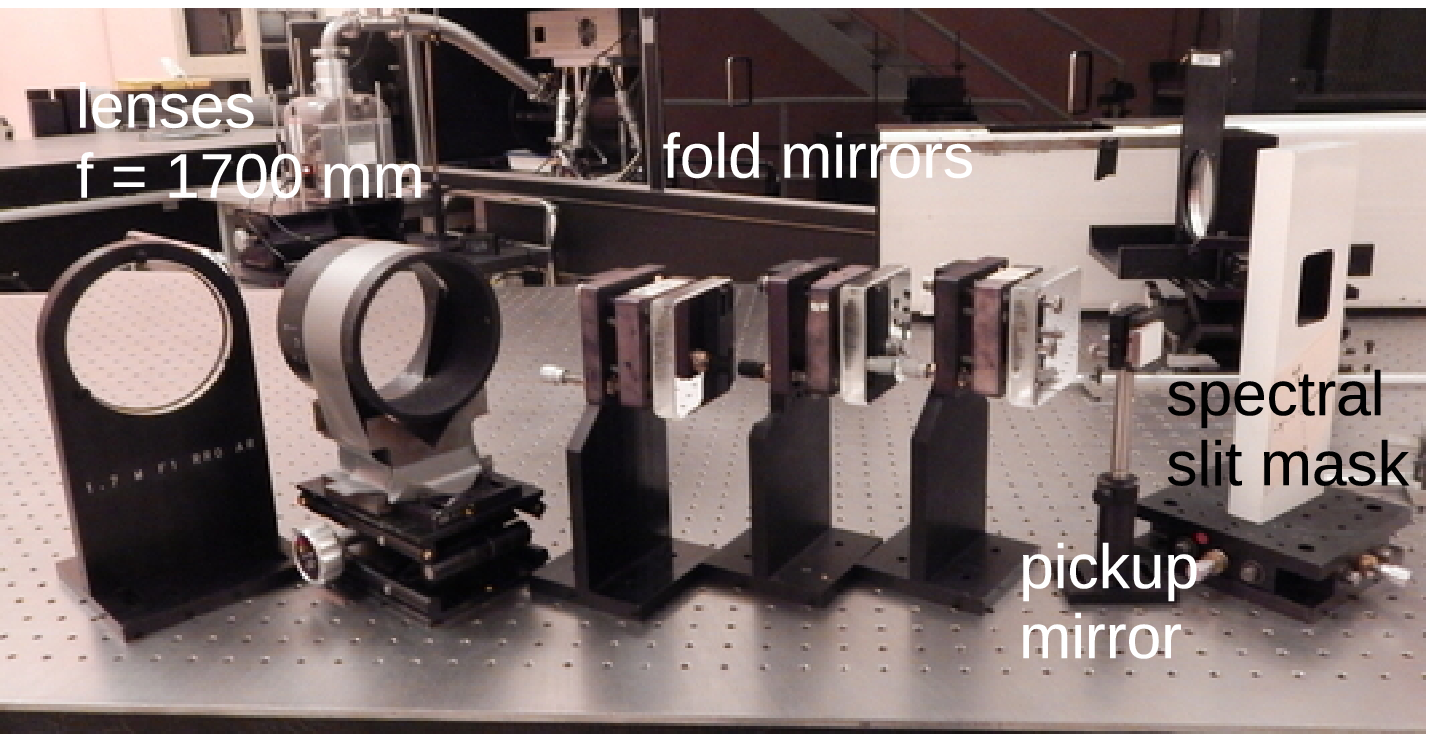}}}
\resizebox{6cm}{!}{\includegraphics{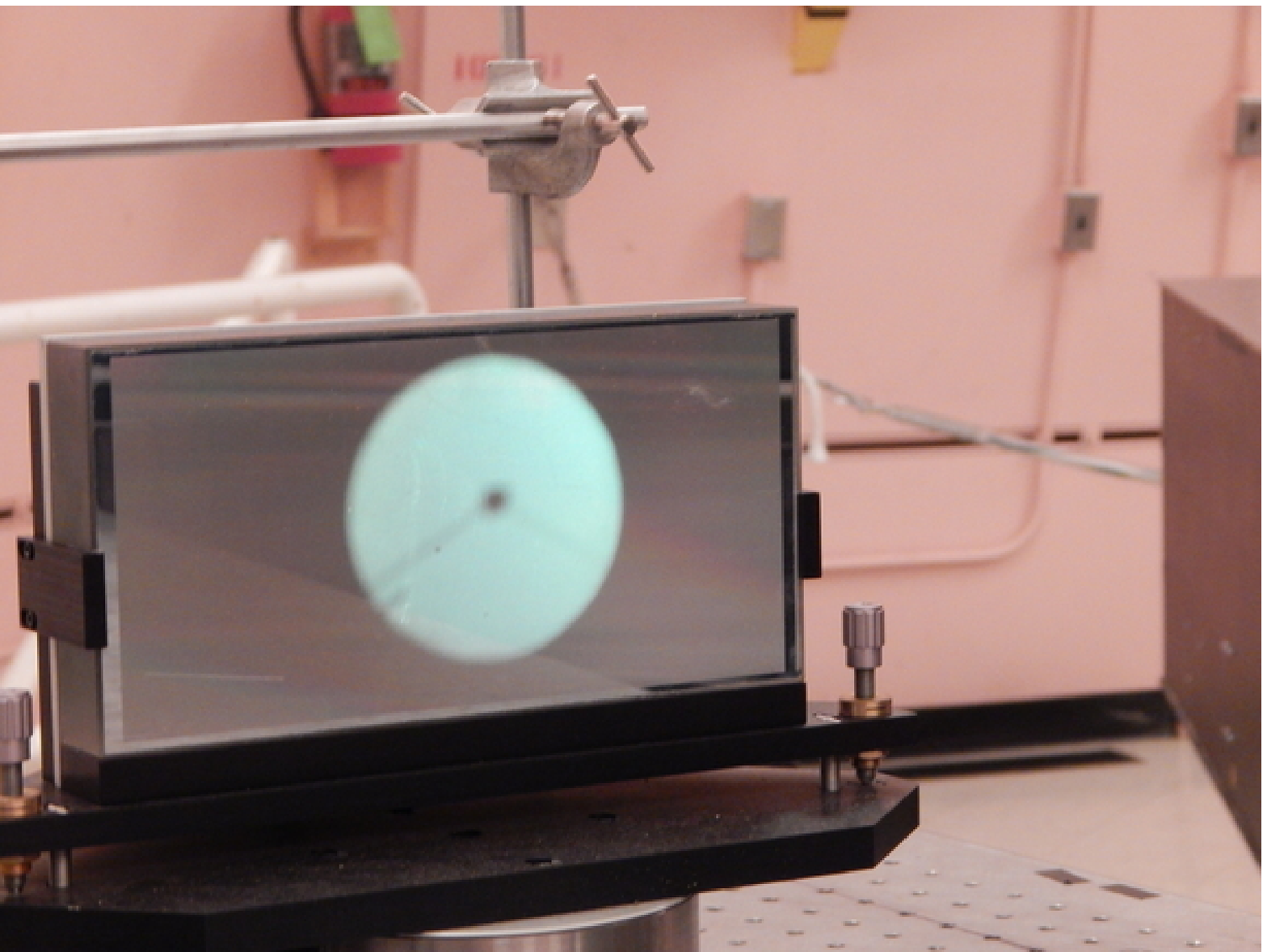}}\hspace*{.5cm}\resizebox{6cm}{!}{\includegraphics{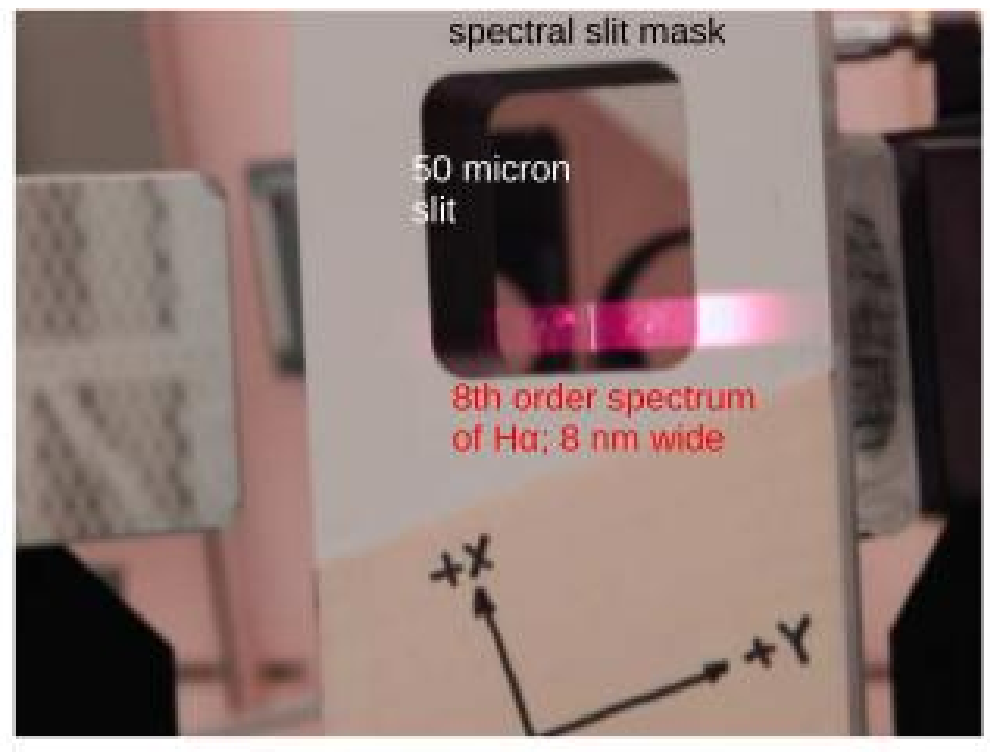}}
\caption{Details of the SDP-internal setup. Top panel: optical elements used. Left to right: camera/collimator lenses with f=1700\,mm, three medium-sized flat fold mirrors, one small pick-up mirror and the 50\,$\mu$m-wide spectral slit mask. Left bottom: white-light pupil image of the first pass on the HSG grating at zero-order orientation. Right bottom: close-up of the spectral slit mask with the dispersed H$\alpha$ spectrum.}\label{setup_details}
\end{figure}
The top panel of Figure \ref{setup_details} shows the optical components other than the grating that were needed for the SDP setup. The two 1.7-m lenses are part of the standard HSG inventory, as is one of the fold mirrors that is always used in the HSG as first optical element behind the slit. For the SDP, two more flat, medium-sized fold mirrors were required to fold the light back to the grating. The pick-up mirror for deflecting the second pass to the side was selected to be as small as possible to prevent vignetting of the incoming beam. The 50\,$\mu$m spectral slit mask was a leftover from another instrument, but we note that we first also tried a 30\,$\mu$m slit from the FIRS instrument and that we also could have used the standard HSG slit mask albeit at some more effort. 

The bottom left panel of Figure \ref{setup_details} shows the problems caused by the fact that the HSG grating was only designed for regular spectrograph observations. The pupil image fills the grating in $y$ to the largest extent. With the grating at 59\,deg close to its blaze angle, the orientation needed to create an H$\alpha$ spectrum in the 8th order, the pupil image actually also fills the grating in $x$. The pupil of the second pass was displaced about 1--2 cm to the right of the incoming pupil, which caused a vignetting of the return beam on the grating and presumably also on the collimator where it was, however, difficult to tell.   

The bottom right panel of Figure \ref{setup_details} shows a close-up of the spectral slit mask. It turned out by trial and error that the astigmatism discussed in the next section was somehow related to it because it reduced when the slit mask was tilted relative to the incoming spectrum. We suspect that actually part of the spectrum reflected on the spectral slit plane was ending up at the focal plane F3. We had placed the reflective side of the slit mask towards the incoming beam, whereas for an SDP one actually would prefer no reflection  at that location. 
\begin{figure}
\centerline{\resizebox{7cm}{!}{\includegraphics{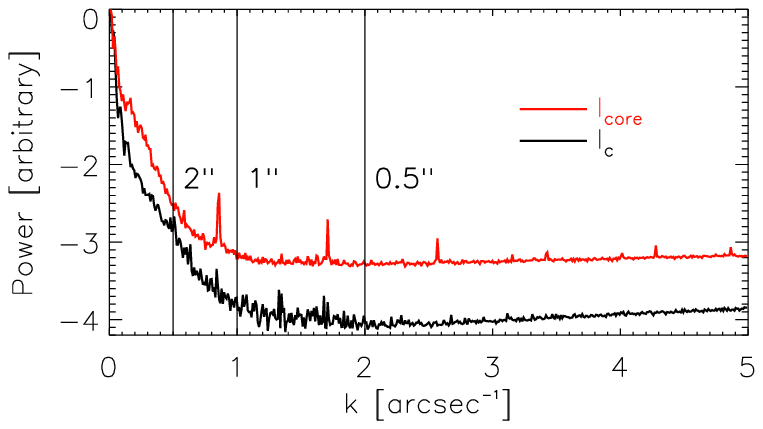}}\resizebox{7cm}{!}{\includegraphics{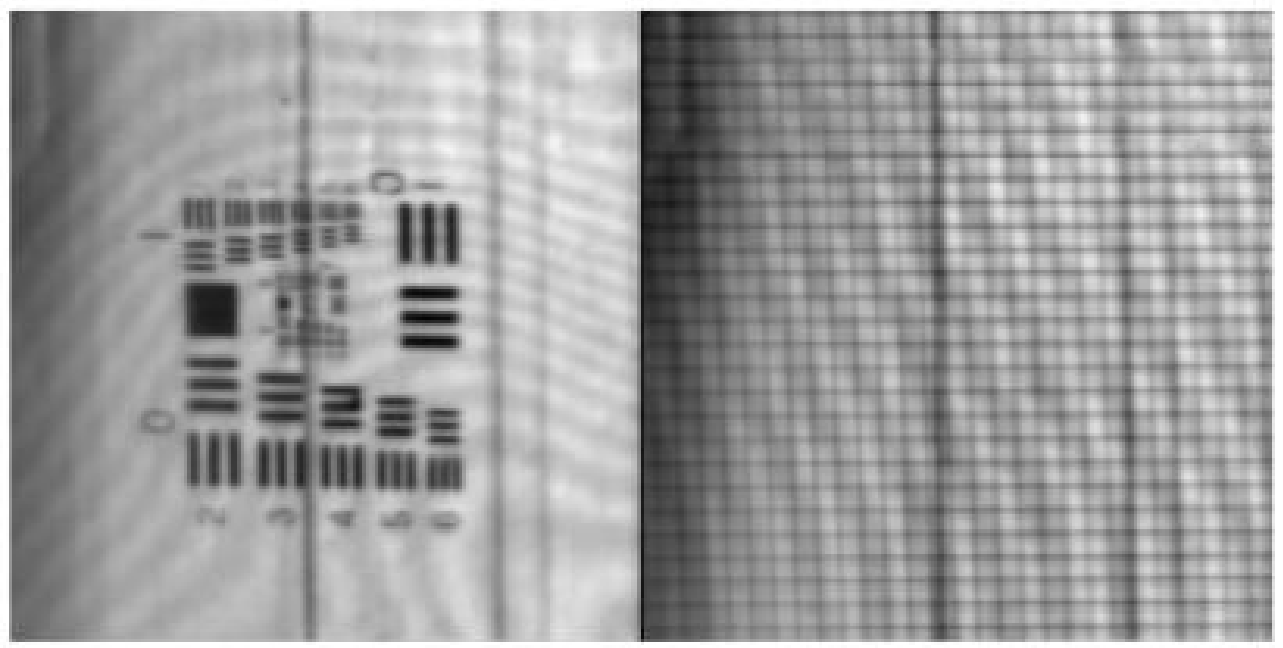}}}
\caption{Spatial resolution of the SDP. Left panel: power spectrum of spatial frequencies for the continuum intensity (black line) and the line-core intensity (red line) of the sunspot observation shown in Figure \ref{burst2exam}.  Right two panels: SDP images of the resolution target (left) and the line grid (right). }\label{sdp_resolution}
\end{figure}   
\subsection{Spatial Resolution}\label{spatres}
\begin{figure}
\centerline{\resizebox{9cm}{!}{\includegraphics{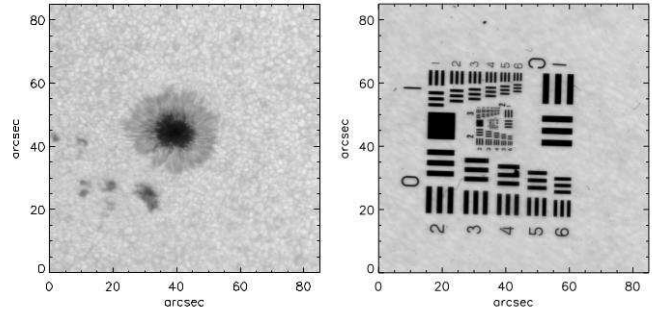}}}
\caption{Example images of the ``slit-jaw'' camera. Left: sunspot image taken on 13 July 2015 at UT 15:57. Right: resolution target. The astigmatism seen in the corresponding SDP image in Figure \ref{sdp_resolution} is missing.}\label{sj} 
\end{figure}

To estimate the spatial resolution, we calculated the Fourier power spectrum of spatial frequencies as described in, \textit{e.g.}, \citet{beck+etal2007}, \citet{puschmann+beck2011} or \citet{katsukawa+etal2012} for the continuum and line-core intensity images of the sunspot shown in Figure \ref{burst2exam}. The images were sampled with 0\farcs1 per pixel corresponding to about critical sampling of the diffraction limit of the DST at 656\,nm. The Fourier power levels off to the constant value that represents the noise level close to, but below 1$^{\prime\prime}$ (left panel of Figure \ref{sdp_resolution}). The corresponding ``SJ'' image (left panel of Figure \ref{sj}) shows that the spatial resolution of the incoming image during the observations was better than the SDP output. A comparison of the resolution target and line grid images taken with the SDP and the SJ (right panels of Figures \ref{sdp_resolution} and \ref{sj}) reveals that the degradation of the spatial resolution happened somewhere inside the SDP part. The SDP images show a clear astigmatism with a worse focus in $x$ than in $y$. Apart from the effect of the orientation of the spectral slit mask on the astigmatism, the vignetting of the return beam on the grating or the slight tilt and off-center position of the second pass on the collimator could have contributed. We would expect that this effect could be made to disappear completely for an oversized grating and collimator and a non-reflective spectral slit mask.
\section{``Slit-Jaw'' Imager}\label{sjimager}
The back-reflection of the order-sorting interference filter at the HSG entrance was fed into the equivalent of a conventional ``slit-jaw'' imager. We used an identical 1k x 1k camera as for the SDP itself with a G-band prefilter. The exposure time for this camera was 35\,ms. Even if the short exposure time also helped to improve the spatial resolution during the observations, the right panel of Figure \ref{sj} with the resolution target demonstrates that also the general focus at the HSG entrance was better than in the final SDP focal plane. 

The SDP images (\textit{e.g.}, Figure~\ref{sdpworks}) themselves also already serve as a sort of SJ image that allows one to optimize, \textit{e.g.}, the telescope pointing even in the absence of a dedicated SJ system, because they contain a combination of spatial and spectral information. The latter allows one to check the spectral range of the observations with one look as well.
 \begin{figure}
\centerline{\resizebox{15.cm}{!}{\includegraphics{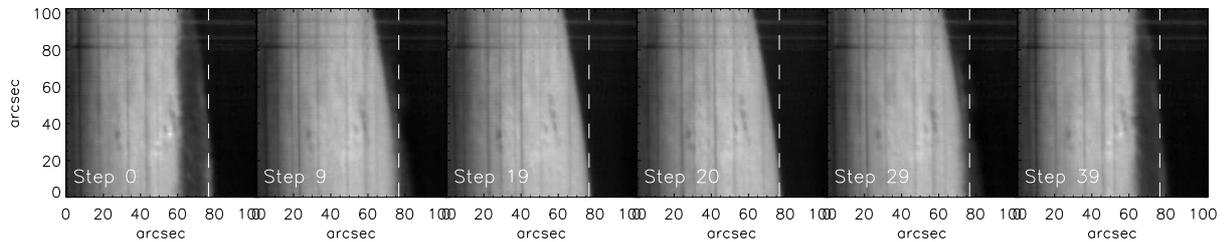}}}$ $\\
\caption{SDP observations of a restricted FOV near the limb. The H$\alpha$ line-core was moved only across about 15$^{\prime\prime}$ in each spectral scan, starting on the disc just off the solar limb. The spectral scanning direction was reversed every 20 steps. The vertical dashed line indicates the location of the spectra shown in Figure \ref{spicules1}. }\label{spicules} 
\end{figure}   
\begin{figure}
\centerline{\resizebox{10.cm}{!}{\includegraphics{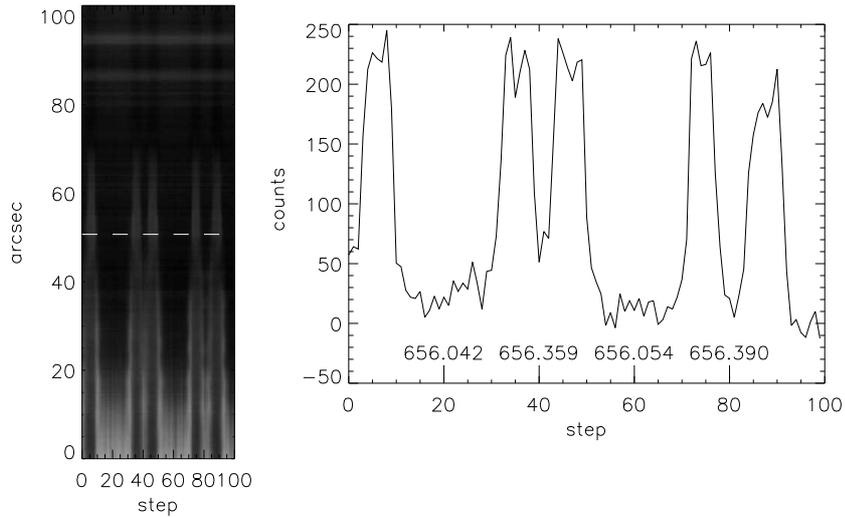}}}$ $\\
\caption{Off-limb spectra close to the solar limb on the column marked in Figure \ref{spicules}. Left panel: spectra along the vertical cuts indicated in Figure \ref{spicules}. Right panel: line profiles at the location of the horizontal dashed line in the left panel. The wavelengths corresponding to four of the steps are denoted at the bottom. }\label{spicules1} 
\end{figure}   
\section{``Fast'' Limb Observations}\label{highcadence}
One option to increase the cadence with an SDP system is to sacrifice spatial coverage. If the spectral range of interest is only moved across a part of the FOV, that region can be observed at a faster cadence while other regions in the FOV in that case spectrally only sample wavelengths that are not deemed interesting. Figures \ref{spicules} and \ref{spicules1} show an example of a SDP observation that was focused only on the limb region to trace spicules. With a restriction of the grating scan to only 20 steps, the H$\alpha$ line core could be moved completely over a 15$^{\prime\prime}$ area from the limb outwards. The time needed for one spectral scan reduced to about 50\,s as this observation was still done with 2.5\,s delay between exposures. The cadence of the spectra depends on the exact step of the reversal of the scanning direction and the column in the FOV selected. We note that there is, however, no restriction of the FOV along the $y$-axis, \textit{i.e.} a $15^{\prime\prime} \times 100^{\prime\prime}$ FOV was scanned in that time frame. 
\end{appendix}
\end{article} 
\end{document}